\documentclass[usenatbib,iop,apj,numberedappendix]{aastex}

\usepackage{amsmath}
\usepackage[normalem]{ulem}
\usepackage{color}

\newcommand{\rate}{\ensuremath{\mathcal{R}}}
\newcommand{\usp}{\ensuremath{u_{\rm sp}}}
\newcommand{\uspfid}{\ensuremath{u_{\rm sp,fid}}}
\newcommand{\uCJ}{\ensuremath{u_{\rm CJ}}}

\newcommand{\To}{\ensuremath{T_{\rm m}}}
\newcommand{\Tb}{\ensuremath{T_{\rm b}}}
\newcommand{\XC}{\ensuremath{X_{\rm C}}}

\begin{document}
\label{firstpage}

\title{A Semi-analytic Criterion for the Spontaneous Initiation of Carbon Detonations in White Dwarfs} 
\author{Uma Garg\altaffilmark{1} and Philip Chang\altaffilmark{1,2}}
\altaffiltext{1}{Department of Physics, University of Wisconsin-Milwaukee, 3135 North Maryland Avenue, Milwaukee, WI 53211, USA; umagarg@uwm.edu, chang65@uwm.edu}
\altaffiltext{2}{corresponding author}
\begin{abstract}
Despite over forty years of active research, the nature of the white dwarf progenitors of Type Ia supernovae remains unclear. However, in the last decade, various progenitor scenarios have highlighted the need for detonations to be the primary mechanism by which these white dwarfs are consumed, but it is unclear how these detonations are triggered.  In this paper we study how detonations are spontaneously initiated due to temperature inhomogeneities, e.g., hotspots, in burning nuclear fuel in a simplified physical scenario. Following the earlier work by Zel'Dovich, we describe the physics of detonation initiation in terms of the comparison between the spontaneous wave speed and the Chapman-Jouguet speed.  We develop an analytic expression for the spontaneous wave speed and utilize it to determine a semi-analytic criterion for the minimum size of a hotspot with a linear temperature gradient between a peak and base temperature for which detonations in burning carbon-oxygen material can occur. Our results suggest that spontaneous detonations may easily form under a diverse range of conditions, likely allowing a number of progenitor scenarios to initiate detonations that burn up the star.
 \end{abstract}

\section{Introduction}\label{introduction}
\noindent
\mathchardef\mhyphen="2D
Type Ia Supernovae (SN Ia) are believed to be the thermonuclear explosion of carbon-oxygen white dwarfs (COWDs), but the nature of the progenitors remains unclear.  Possible progenitors include COWDs near the Chandrasekhar-limit of $M_{\rm Ch} = 1.4\,{\rm M}_{\odot}$ (\citealt{1973ApJ...186.1007W}; \citealt{Nomoto+86}, for more modern reviews see \citealt{2013FrPhy...8..116H} and \citealt{Maoz+14}), mergers of WDs  \citep{2009A&A...500.1193L,van_Kerkwijk+10,2012ApJ...747L..10P,Zhu+13,2014MNRAS.438...14D}, detonations of an outer He layer that trigger a central detonation \citep{Livne+90, Fink+07, 2012MNRAS.420.3003S,Ken+14}, collisions of white dwarfs \citep{Raskin+09,2012ApJ...747L..10P, Kushnir+14}, and core explosions of massive stars \citep{Kashi+11}.  

These progenitor scenarios for SN Ia employ detonations as the mechanism by which CO is burnt.  While it was initially thought that deflagration is the mechanism by which these WDs explode \citep{Nomoto+86}, state-of-the-art calculations of deflagration burning in near-Chandrasekhar mass COWDs showed that the resulting explosions are weak and may result in bound remnants or contaminated with a sizeable amount of unburnt material, which is not seen observationally \citep{Maoz+14,2014MNRAS.438.1762F}

Detonations offer a way to overcome these problems. In particular, a detonation at an appropriate density will burn the star completely and produces the correct observed elemental abundances \citep[see for instance ][]{Sim+10}. Hence, for near-Chandrasekhar mass COWDs, it is generally thought that an initial deflagration transitions to a detonation, i.e., a deflagration-to-detonation transition (DDT; \citealt{Woosley90,Khokhlov+97,Niemeyer+97,2000ApJ...538..831L,2007ApJ...668.1103R,2007ApJ...668.1109W,Seitenzahl+09}, hereafter S09, \citealt{2013MNRAS.429.1156S,2013A&A...559A.117C}).  In addition, for sub-Chandrasekhar explosions and for WD collisions, a detonation is also required to burn the WD.

For the colliding WD or He layer detonation scenarios, a detonation results from the impulse imparted by the collision or the geometric focusing of a shock wave, i.e. a prompt detonation.  For the other scenarios, i.e., DDT and sub-$M_{\rm Ch}$ mergers, it is thought that detonations are triggered spontaneously.  Spontaneous detonations may occur when burning material develops temperature inhomogeneities, i.e., hotspots,  that precondition the fuel.  These hotspots may develop from the interaction between the deflagration flame and turbulence as in DDT \citep{Khokhlov+97,2000ApJ...538..831L,2007ApJ...668.1103R,2007ApJ...668.1109W} or from the endpoint of the ``simmering phase'' of a sub-Chandrasekhar mass WD (Chang et al., in preparation). 

The triggering of detonations is also of active interest in a more terrestrial context. \citet{Zel'dovich+70} performed some of the first numerical experiments on spontaneous detonations in linear temperature gradients.  Here, he showed that detonations occur for temperature gradients that are neither too shallow nor too steep.  In particular, \citet{Zel'dovich+70} parameterized the classification of shallow and steep gradient in terms of the comparison between the spontaneous wave speed, $\usp$, the speed at which a thermal runaway moves down a temperature gradient, and the Chapman-Jouguet (CJ) speed, $\uCJ$, the speed of self-sustaining detonation.  In the WD context, previous  works including \citet{1994ApJ...427..330A},\citet{Niemeyer+97}, \citet{Dursi+06}, \citet{2007ApJ...660.1344R}, and S09 utilized similar numerical experiments to determine the steepest temperature gradient, and, hence, the smallest hotspot size, i.e., the critical radius, that would allow for spontaneous detonations.   However, the relation between $\usp$ and $\uCJ$ has not been actively explored  in the previous literature. 

Motivated by this problem, we explore in this paper how detonations are spontaneously triggered inside hotspots and how the success or failure of detonation initiation depends on the spontaneous wave speed and CJ speed.  In Section \ref{The Physics of Detonation Initiation}, we discuss the physics of detonations and their spontaneous initiation in non uniform temperature regions known as hotspots.  We also develop a semi-analytic condition for determining the critical hotspot radii that lead to successful detonations in Section \ref{sec:semianalytic}. Finally, we summarize, compare with previous work, and conclude in Section \ref{Conclusions}.

\section{Induction time and Spontaneous wave speed}\label{The Physics of Detonation Initiation}

Motivated by the importance of detonations for SN Ia, we study how detonations are initiated in material undergoing nuclear burning in this work.
We begin by discussing the heating of CO material undergoing nuclear burning. We consider temperatures of $T < 2 \times 10^9$ K typical of the end of simmering where C-C reactions dominate energy generation.
For our calculations, the following C-C nuclear reactions were considered :
\begin{equation}\label{eq:Na}
^{12} \text{C} + ^{12} \! \text{C}  \rightarrow  p + ^{23}  \! \text{Na}
\end{equation}
\begin{equation}\label{equation4}
^{12} \text{C} + ^{12} \! \text{C}  \rightarrow  n  + ^{23}  \! \text{Mg}
\end{equation}
\begin{equation}\label{eq:Ne}
^{12} \text{C} + ^{12} \! \text{C}  \rightarrow  ^{4} \! \text{He} + ^{20}  \! \text{Ne}
\end{equation}
We consider the hydrodynamic equations, which are the continuity equation, 
\begin{equation}\label{equationa}
\frac{\partial \rho}{\partial t} + \frac{\partial (\rho u)}{\partial x}= 0,
\end{equation}
the momentum equation,
\begin{equation}\label{equationb}
\frac{\partial (\rho u)}{\partial t} + \frac{\partial (P + \rho u^2)}{\partial x}= 0,
\end{equation}
and the energy equation,
\begin{equation}\label{equationc}
\frac{\partial(\rho E)}{\partial t} +\frac{\partial((P+\rho E)u)}{\partial x}= \dot{Q},
\end{equation}
where $\rho$ is density, $u$ is the 1-d velocity, i.e., x-velocity, and $P$ is the pressure. 
The energy, $E$, is the sum of the specific internal energy\footnote{We presume that the internal energy is dominated by the non-degenerate ions.  The contribution of the degenerate electrons are suppressed relative to the non-degenerate ions because only their fluctuations above the Fermi sea contribute to the internal energy, and hence it is suppressed by a factor of $k_B T/E_{\rm F}$, where $E_{\rm F}$ is the Fermi energy. } and kinetic energy,
\begin{equation}\label{equation7}
E= \frac{k_B T}{(\gamma - 1)A m_p}+\frac{u^2}{2},
\end{equation}
where $ m_p$ is proton mass, $A$ is mean atomic number of the background material, and $\gamma = \Gamma_3 = 5/3$ is the third adiabatic exponent for the ion gas, which is treated as an ideal gas. Here we assume that the internal energy is supplied by the ions, which is a reasonable approximate for degenerate material as long as the temperature is not too large.  We also note that we ignore conduction as this timescale is typically long compared to the timescale for ignition at these temperatures (see for instance, \citealt{Dursi+06}). The heating rate per unit volume, $\dot{Q}$, is given by,
\begin{equation}\label{equation2} 
 \dot{Q} =\left(\frac{\rho \XC}{A m_p}\right)^2\sum_i Q_i \rate_i 
 \end{equation} 
where $Q_i$ and $\rate_i$ are the energy released per reaction and reaction rate coefficient per reaction $i$, respectively.\footnote{Rates derived from: http://ie.lbl.gov/astro2/rate5.txt}

Below, we study a hotspot with a linear temperature gradient between the peak temperature $\To$ and a base temperature $\Tb$.
As we will be interested in C-C burning near $\To$, we perform a power law expansion on the temperature dependence of the reaction rate coefficient to find
\begin{equation}\label{eq:qdot}
\dot{Q} = \left(\frac{\rho \XC}{A m_p}\right)^2\sum_i Q_i \rate_{i,0} \left(\frac{T}{\To}\right)^{\alpha_i}, 
\end{equation}
where $\rate_{i,0} = \rate_i(T=\To)$ is the reaction rate coefficient at $T=\To$ and $\alpha_i={\partial \ln \rate_i}/{\partial \ln T}$ is the exponent of the power law dependence on temperature evaluated at $T = \To$.  In Table \ref{table:1.6}, we tabulate values of  $\alpha_i$, $Q_i$\footnote{Values of $Q_i$ taken from \citet{Clayton68}} and  $\rate_{i,0}$ for the three C-C reactions (eqns [\ref{eq:Na}], [\ref{equation4}], and [\ref{eq:Ne}]). Since, the Mg reaction (eq.[\ref{equation4}]) is slow, we neglect this reaction in our calculations below.  Noting that the two other reactions (eqns [\ref{eq:Na}] and [\ref{eq:Ne}]) have nearly the same dependence on temperature, i.e., $\alpha_i \approx \alpha = 22$, we can apply an appropriate sum over the reaction rates, namely
\begin{equation}\label{eq:energy rate}
\sum_i \rate_i Q_i = \left(Q_{\rm Na} \rate_{\rm Na, 0} + Q_{\rm Ne} \rate_{\rm Ne, 0}\right)\left(\frac T {\To}\right)^{\alpha}.
\end{equation}

\begin{table}
\caption{\label{table:1.6} Values  of  $\alpha_i$  and  $\rate_{i,0}$ at $\To = 1.6 \times 10^9\,{\rm K} $}
\centering
\begin{tabular}{c c c c}
\hline
Reaction name & $\alpha_i$ & $\rate_{i,0}\tablenotemark{a} $ & $Q_i$ (MeV)\\
\hline
$^{12} \text{C} + ^{12} \! \text{C}  \rightarrow  p + ^{23}  \! \text{Na}$ &  22.31 & $2.62\times10^{-30}$ & 2.24\\
\hline
$^{12} \text{C} + ^{12} \! \text{C}  \rightarrow  ^{4} \! \text{He} + ^{20}  \! \text{Ne}$ &  22.1861 & $3.12\times 10^{-30}$ & 4.62\\
\hline
$^{12} \text{C} + ^{12} \! \text{C}  \rightarrow  n  + ^{23}  \! \text{Mg}$ &  31.4099 & $ 9.33\times 10^{-32}$ & -2.60\\
\hline
\end{tabular}
\tablenotetext{1}{Units of $\rate$ are in cm$^3$ s$^{-1}$}
\end{table}

When hydrodynamics is suppressed, i.e. $u = 0$, evolution happens only in time and $x$ appears as a parameter through the initial conditions. The hydrodynamic equations then reduce to:
\begin{equation}
\frac{1}{\gamma - 1}\frac{\partial P}{\partial t} = \dot{Q},
\end{equation}
\begin{equation}\label{equation1}
\frac{\partial T}{\partial t} = \frac{(\gamma -1)A  m_p \dot{Q}}{\rho k_B}.
\end{equation}
\noindent
The initial conditions are uniform except for a linear gradient in temperature given by,
\begin{equation}\label{equation3}
T(x,0)= \To\left(1- \frac {\Delta T}{\To}\frac x {\lambda}\right), 
\end{equation}
where, $\Delta T = \To - \Tb$ and $\lambda$ is the length of the region over which $T$ varies. 
Equation (\ref{equation1}) can now be solved with the approximation that $X_C$, $\rho$, and $A$ remained fixed in time and that the burning is due completely to C-C reactions that is given by equation (\ref{eq:qdot}) and using the approximation of equation (\ref{eq:energy rate}).  While this is not true in reality, these assumptions are valid for the initial burning when only a small fraction of the initial material is consumed. Toward that end, Equation (\ref{equation1}) can written as, using equations (\ref{eq:qdot}) and (\ref{eq:energy rate}):
\begin{equation}\label{eq:on the way}
\frac{1}{T_m}\frac{\partial T}{\partial t} = \frac{(\gamma -1)\rho \XC^2\left(Q_{\rm Na} \rate_{\rm Na, 0} + Q_{\rm Ne} \rate_{\rm Ne, 0}\right)}{A m_p k_B T_m}\left(\frac T {\To}\right)^{\alpha}.
\end{equation}
This is now integrated for every point x using the initial conditions given by equation (\ref{equation3}) to find:
\begin{equation}\label{temp}
T = T(x,0) {\left(1-\frac{t}{\tau_{i}}\right)}^{\frac{1}{1- \alpha}},
\end{equation}
where $\tau_{i}$ is the induction time, the time at which the nuclear fuel burns out completely, and the total heat released $Q$ can be written as:
\begin{equation}
Q = \int_{0}^{\tau_i}\dot{Q} dt.
\end{equation}
Using equation (\ref{equation2}),  $\tau_i$ is then given by:
\begin{equation}\label{tau}
{\tau}_i = {\tau}_{i,0} {\left[1-\left(\frac{\Delta T}{\To \lambda}\right)x\right]}^{(1-\alpha)} 
\end{equation}
where,
\begin{equation}\label{tau1}
{\tau}_{i,0} = \frac{A m_p k_B \To}{(\alpha - 1)(\gamma -1) \rho {\XC}^2 \left(Q_{\rm Na} \rate_{\rm Na, 0} + Q_{\rm Ne} \rate_{\rm Ne, 0}\right)}
\end{equation}
is the induction time at the center of the hotspot with temperature $T=\To$.

\begin{figure*}
\includegraphics[width=0.48\textwidth]{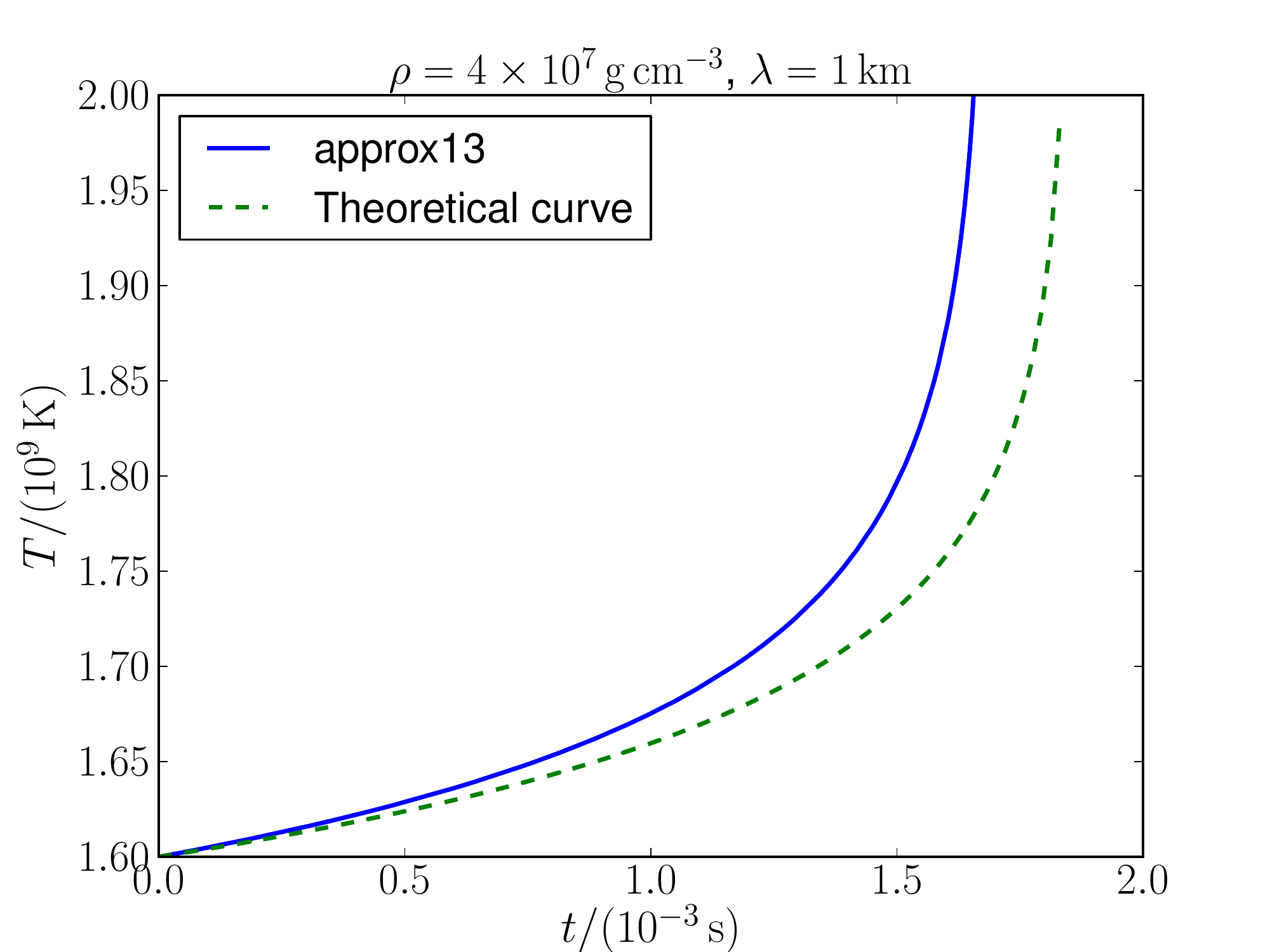}
\includegraphics[width=0.48\textwidth]{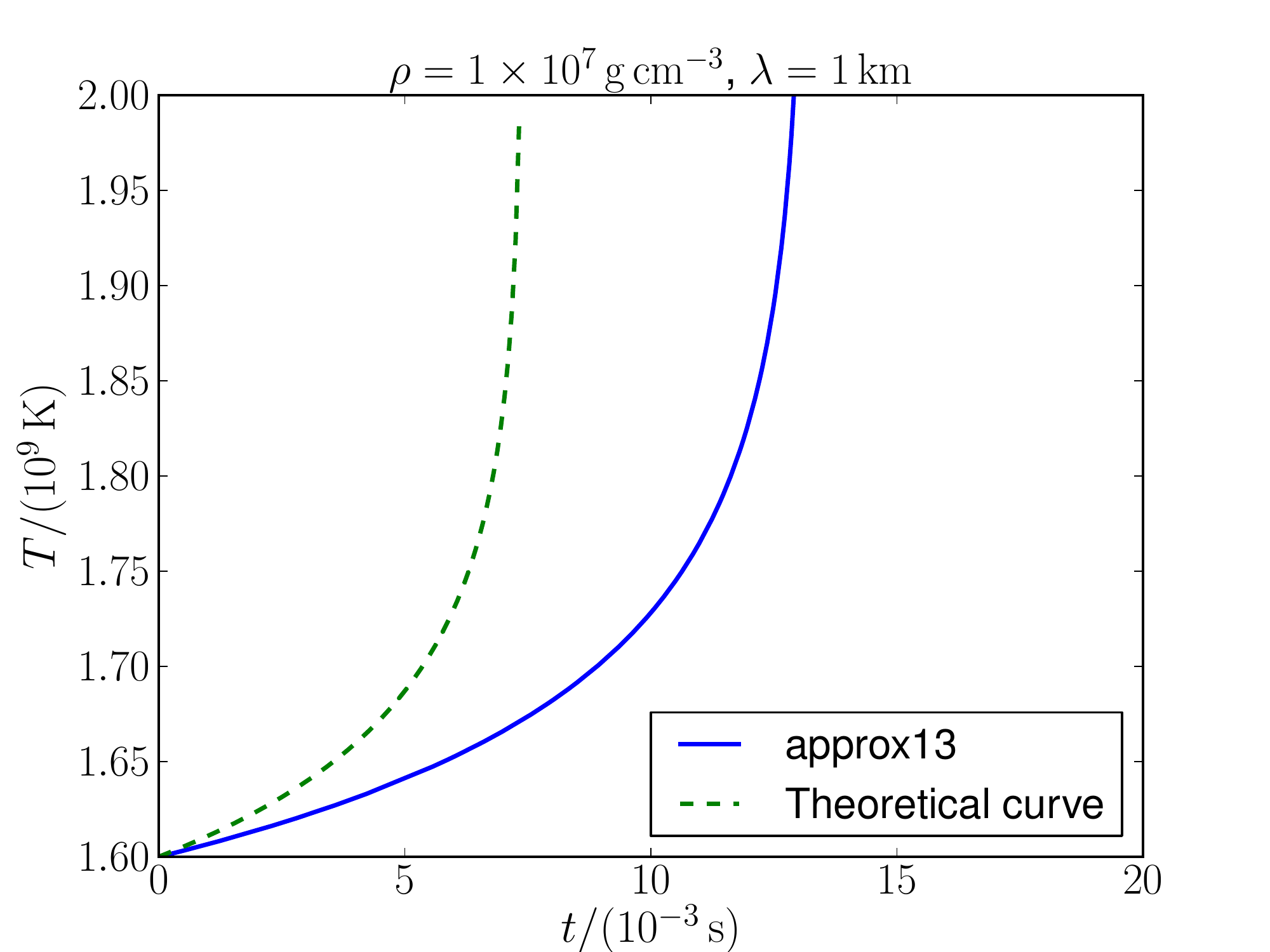}

\caption{Evolution of temperature with respect to time for $\rho = 4\times 10^7\,{\rm g\,cm}^{-3}$ (left) and $\rho = 10^7\,{\rm g\,cm}^{-3}$ (right) for an initial temperature of $T=1.6\times 10^9$ K. In both plots, as $t \rightarrow \tau_{i}$, the temperature increases steeply, which leads to thermal runaway. The agreement for $\tau_i$ between the theoretical curve, calculated using equation (\ref{temp}) and the approx13 curves, calculated numerically using the approx13 nuclear network, is within a factor of two.\label{fig:induction} }
\end{figure*}

Formally, equation (\ref{temp}) implieds that as $t\rightarrow \tau_i$, the temperature runs away to infinity.  In reality this does not happen as the assumption of constant $X_C$ breaks down as the carbon is rapidly consumed by nuclear burning as $t\rightarrow \tau_i$.  In reality, the temperature will run away and approach the maximum temperature that can be reached from consumption of all the available nuclear fuel.  In Figure \ref{fig:induction}, we plot the temperature at the center of the hotspot as a function of time, $t$,  for $\rho$ = 4 $\times 10^7 {\rm g\,cm^{-3}}$ (top plot) and $\rho$ = 1.0 $\times 10^7 {\rm g\,cm^{-3}}$ (bottom plot), calculated using equation (\ref{temp}). We also integrate equation (\ref{equation1}) numerically using the approx13 \citep{Timmes99} nuclear burning network (solid lines).\footnote{The approx13 nuclear network is available from \url{http://cococubed.asu.edu/}.}
In both cases, as $t \rightarrow \tau_i$, the temperature, $T$, increases steeply as the nuclear fuel burns. The induction time calculated analytically agrees with the numerical result to within a factor of two for both cases.  Note here that we restrict the temperature range to preserve the assumption of constant $X_C$ in comparing the approximate curve given by equation (\ref{equation1}) to the full numerical result.

Because $\tau_i$ is a function of the initial temperature, it varies along a temperature gradient.  Namely, initially hotter regions run away first, followed by cooler regions.  This allows us to define a speed known as the spontaneous wave speed, $\usp$, which is the speed at which a thermal runaway moves down the temperature gradient when hydrodynamics is suppressed.  Using equation (\ref{equation3}), we can rigorously define this as (\citealt{2002CTM.....6..553K}, S09):
\begin{equation}\label{spontaneous}
\usp  = \left(\frac{d\tau_i}{dx}\right)^{-1} =  - \left(\frac{d\tau_i}{dT}\right)^{-1} \left(\frac{\lambda}{\Delta T}\right). 
\end{equation}
This velocity can be compared to the Chapman-Jouguet (CJ) speed, which is the speed of a self-sustaining detonation and is given by \citep{Ken+14} 
\begin{equation}\label{eq:CJ speed}
\uCJ = \sqrt{2(\Gamma_1^2-1)Q_{\rm det}},
\end{equation}
where $\Gamma_1 = (\partial \ln P/\partial \ln \rho)_S$ is the first adiabatic index, which for relativistic, degenerate gas is $4/3$, and $Q_{\rm det}$ is the total energy released per gram by the detonation.  

A comparison between the spontaneous wave speed, $\usp$, and the CJ speed, $\uCJ$ forms the basis of the so-called Zel'Dovich criterion \citep{Zel'dovich+80}. In particular, when $\usp $  is greater than $\uCJ$, detonations do not occur, i.e., as $\usp $  approaches infinity, a constant volume explosion takes place. When $\usp \approx \uCJ$, then a detonation may initiate.  Finally for $\usp \ll \uCJ$, a detonation is not triggered.  This last limit is especially interesting as it defines the smallest hotspot region that undergoes detonation, i.e., the critical radius.  In this limit, it is thought that detonations are triggered by the so-called SWACER mechanism (see the discussion in S09).  We note, however, that the SWACER mechanism is not well described analytically and thus a simple criterion for determining the critical radius does not exist in the literature.

\section{Semi-analytic condition for detonations}\label{sec:semianalytic}

In an effort to develop a semi-analytic criterion for the critical radius, we will use extensive critical radii tabulated by S09 and take advantage of the analytic results for the induction time that we had developed in Section \ref{The Physics of Detonation Initiation}. Using equations (\ref{tau}), (\ref{tau1}), and (\ref{spontaneous}), we find:
\begin{equation}\label{usp}
\usp(x) = {\left[1-\frac{\Delta T}{\To}\frac x { \lambda}\right]}^\alpha \left(\frac{\To \lambda}{\Delta T (\alpha-1) \tau_{i,0}}\right).
\end{equation}
To determine $\tau_{i,0}$, we take equation (\ref{tau1}) to find:
\begin{equation}
\tau_{i,0} = \Theta(\To) X_{C,0.5}^{-2}  \rho_7^{-1}\,{\rm s},\label{inductime}
\end{equation}
where $X_{\rm C,0.5} = \XC/0.5$, $\rho_7 = \rho/10^7\,{\rm g\,cm}^{-3}$, and $\Theta$ is the normalization that is determined completely by $\To$.  
Here to extend our results to the larger temperature ranges covered by S09's parameter study, we calculate and tabulate $\alpha$ and $\Theta$ for the various $\To$ studied by S09 and ourselves in Table \ref{table:alphas and taus}.
\begin{table}[t]
\caption{\label{table:alphas and taus} Values  of  $\alpha$  and  $\Theta$ for different $\To$'s}
\centering
\begin{tabular}{c c c}
\hline
$\To$ & $\Theta$ & $\alpha$\\
\hline
$1.6\times 10^9$ K & $8.9\times 10^{-3}$ & $22$ \\
$1.8\times 10^9$ K &  $8.1\times 10^{-4}$ & $21$ \\
$2.0\times 10^9$ K &  $1.0 \times 10^{-4}$ & $20.3$ \\
$2.4\times 10^9$ K &  $3.4\times 10^{-6}$ & $18.8$ \\
$2.8\times 10^9$ K &  $2.7\times 10^{-7}$ & $17.7$ \\
\hline
\end{tabular}
\end{table}

The induction time given by equation (\ref{inductime}) can be compared to other analytic expressions of the induction time given by \citet{Dursi+06} and \citet{Woosley+04}.  Our derivation of the induction time is most similar that of \citet{Woosley+04} with the crucial difference that we ignore the electron screening correction which is unimportant for the densities that we are concerned about, but are important for the high densities that \citet{Woosley+04} studied.  Thus, the results of \citet{Woosley+04} cannot be compared to our results due their assumption of the dominance of the screening correction.  A more direct comparison can be made between the analytic formula of \citet{Dursi+06} (their equation (2)) and equation (\ref{inductime}).  Here, we find that the induction time that we compute compared against the induction time fitted by \citet{Dursi+06} varies by an order of magnitude.  Part of the discrepancy may come from the fact that the fit given by \citet{Dursi+06} is good only to a factor of five compared to the numerically computed results.  
Using equation (2) of \citet{Dursi+06}, we find $\tau_i(T_9=1.6) \approx 0.033$ s, while we find $\tau_i = 0.0089$, a factor of three difference.  The numerically computed result, which can be see in the lower plot of Figure \ref{fig:induction}, suggests that the $\tau_i \approx 0.013$ s, which can be seen from where the approx13 curve becomes nearly vertical, is between these two values. Hence, we note that our estimate for the induction time may be significantly different from that of \citet{Dursi+06}, but it is within the range of their fit, and both estimates for $\tau_i$ is within an order of magnitude of the numerically computed $\tau_i$.

Returning to equation (\ref{usp}), we note that $\usp$ is a function of position, which implies that it depends on the background or base temperature in which a hotspot sits.  Namely, the lower the base temperature, the lower the minimum $\usp$.  However, as our discussion above suggests, for low $\usp$, the detonation starts well before the minimum $\usp$ is reached, i.e., it starts close to the center of the hotspot.  This would suggest a successful initiation of a detonation depends on $\To$ and the gradient. 

If we adopt the view that the gradient in induction times is the determining factor in determining whether a detonation starts or fizzles, then we must evaluate $\usp$ as some fiducial position $x_{\rm fid}$, which is the same as evaluating $\usp$ at some fiducial temperature, $T_{\rm fid}$.  This allows us to define some fiducial $\uspfid$ as
\begin{eqnarray}
\uspfid &=& \usp(x_{\rm fid}) = {\left[1-\frac{\Delta T}{\To}\frac {x_{\rm fid}} { \lambda}\right]}^\alpha \left(\frac{\To \lambda}{\Delta T (\alpha-1) \tau_{i,0}}\right)\nonumber\\
&=&  {\left[\frac{T_{\rm fid}}{\To}\right]}^\alpha \left(\frac{\To \lambda}{\Delta T (\alpha-1) \tau_{i,0}}\right),
\label{eq:uspfid}
\end{eqnarray}
where $T_{\rm fid} = \left(\To - \Delta T(x_{\rm fid}/\lambda)\right)$.

To complete our condition for detonation initiation, we must set $\uspfid$ to some speed.  Motivated by \citet{Zel'dovich+70}'s and our own results on the importance of the CJ speed, we choose the condition
\begin{equation}\label{eq:uspfid=betaucj}
\uspfid=\beta\uCJ,
\end{equation}
where $\beta$ is a numerical factor that is less than unity.\footnote{We note that the other possible choice is the sound speed.  For the conditions that we study here, i.e., using our and S09's results, where $\rho = 1-4\times 10^7\,{\rm g\,cm}^{-3}$, there is insufficient dynamic range in the sound speed to determine which a more appropriate choice.}  We estimate $Q_{\rm det}$ for determining $\uCJ$ from equation (\ref{eq:CJ speed}) from the difference of the binding energies of iron-group material, i.e., $^{56}$Ni, and C to find $\uCJ \approx 1.29 \times 10^9\,{\rm cm\, s}^{-1}$; we assume that the material is completely burned in the detonation.  This estimate for \uCJ\ is checked against one dimensional numerical simulations of detonations using FLASH \citep{Fryxell+00}, a massively parallel, adaptive mesh refinement hydrodynamical code. In our simulations, we have set up a 1000 km 1D box with an appropriate temperature gradient in the center that will lead to a detonation and use the AMR capabilities of FLASH to set a minimum resolution of $\approx$ 4 m and a maximum resolution of $\approx $ 0.5 m. A more detailed discussion of these simulations and their results is deferred to future work.  In any case, for the purpose of this work, we have measured the detonation speed after the detonation fully develops and find that it is in excellent agreement with the estimate of $\uCJ$ that we describe above. 

Hence, the condition for critical gradient for detonation reduces down to determining $T_{\rm fid}$ and $\beta$.  To fit these two parameters, we use both the numerical results for this study where $\rho_7=1$ and $4$ and the tabulated results of S09 for linear gradients for $\rho_7 = 1$ (their table 1 for $X_C = 0.5$ and tables 8-11 for $X_C = 0.3-0.7$).  S09 tabulate their result in terms of critical radii.  To convert this to a speed, we use their critical radius to set $\lambda=\lambda_{\rm crit}$ in equation (\ref{eq:uspfid}).  We then make a rather inspired guess for $T_{\rm fid}$ to be
\begin{equation}\label{eq:Tfid}
 T_{\rm fid} = \frac {\To + T_o}{2},
\end{equation}
where $T_o$ is set to be $ 1.5\times 10^9$ K.  

\begin{figure}
\includegraphics[width=0.8\textwidth]{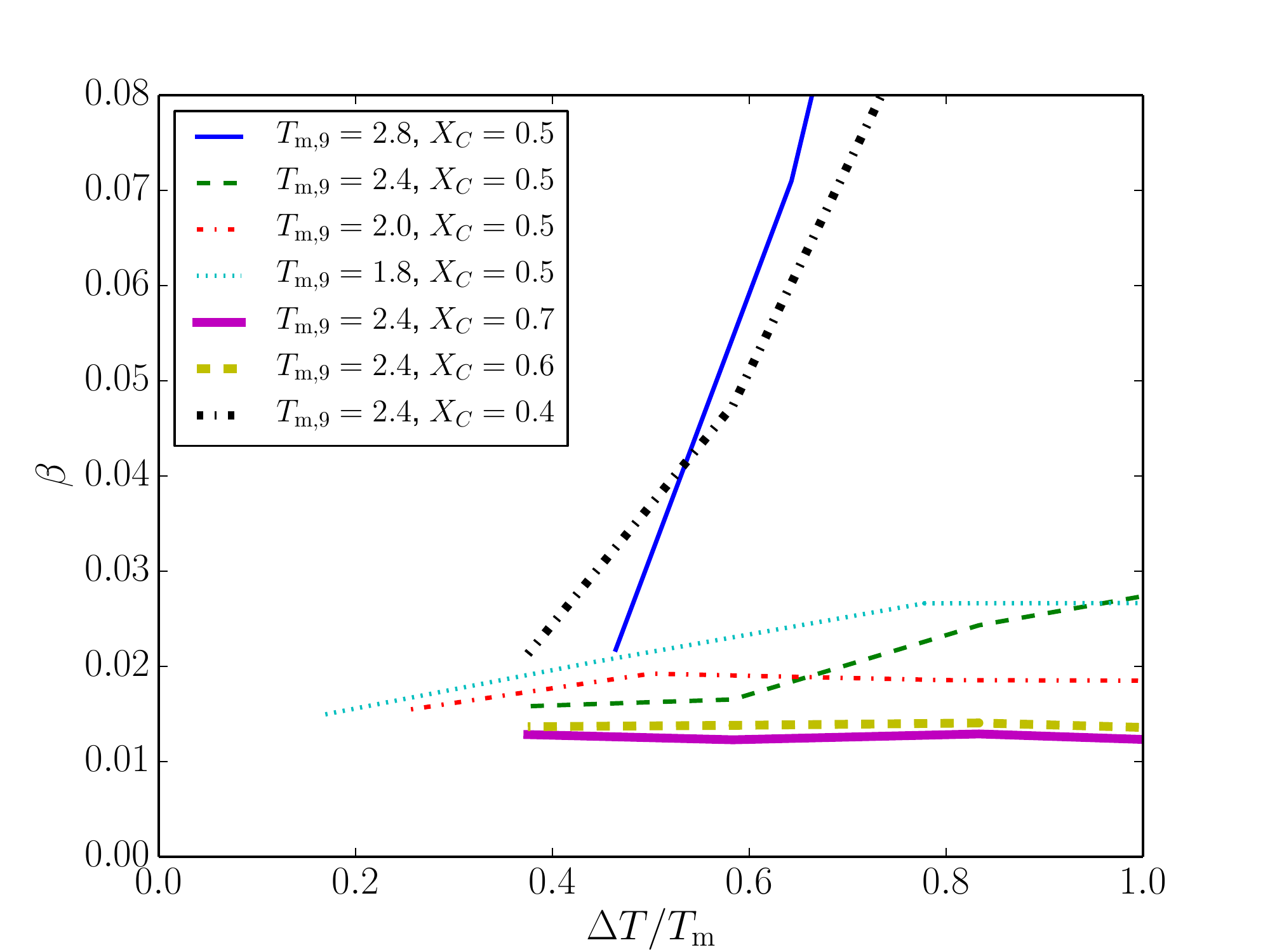}
\caption{Ratio between $\uspfid$ and $\uCJ$ at the critical radii, $\lambda_{\rm crit}$, for the initiation of detonations.  For $X_C \ge 0.5$ and $\To \ge 2.4\times 10^9$ K, $\beta\approx 0.018$ is roughly flat with a variation of about a factor of two.  The flatness of $\beta$ as a function of $\Delta T/\To$ implies that the critical radii scales linearly with $\Delta T/\To$, implying that gradient does not change as function of the base temperature, $T_{\rm b}$.  For $X_C < 0.5$ and $\To > 2.4\times 10^9$ K, the fit breaks down, but this might be a result of the oxygen burning which is not accounted for in our analytic analysis.\label{fig:beta}}
\end{figure}

We plot $\beta = \uspfid(T_{\rm fid})/\uCJ$ as a function of $\Delta T/\To$ in Figure \ref{fig:beta} using the prescription for $T_{\rm fid}$ given by equation (\ref{eq:Tfid}).  It is clear from this figure that $\beta$ is flat and sits in a rather narrow range around $\approx 0.02$ for S09's results for $\To \leq 2.4\times 10^9$ K and $X_C \geq 0.5$.  This is remarkable as the range in critical radii tabulated by S09 vary by two orders of magnitude, yet our physically motivated fit to the critical radii over its entire range vary about a factor of two at most.  The exception is for low carbon fraction ($X_C \leq 0.4$) and high $\To$ ($\To \geq 2.8\times 10^9$ K).  Here we suspect that oxygen burning becomes increasingly important for these parameter regimes for which our pure carbon burning expression for $\usp$ breaks downs.  Lending credence to this suspicion is that for high carbon fraction $X_C = 0.6$ (thick gold dashed line) and $ 0.7$ (thick magenta solid line), the parameter $\beta$ is remarkably flat, following the condition even better than for the case of $X_C = 0.5$.  

Figure \ref{fig:beta} also shows that the critical radii determined by S09 scales linearly with $\Delta T/\To$.  This shows that it is the gradient  that is relevant for the initiation of a detonation, which S09 also realized.  In addition, the critical radii nearly scales like $X_C^{-2}$ for $X_C \geq 0.5$.  Both of these trends are apparent in the S09's tabulated results, as they also noted, but also arise naturally from our expression for the fiducial spontaneous wave speed (eq.[\ref{eq:uspfid}]).   

Fitting for $\beta$ using the results of S09 for $\To \leq 2.4\times 10^9$ K and $X_C \geq 0.5$, we find $\beta \approx 0.018 \pm 0.002$.   Hence, our semi-analytic condition for the critical gradient for spontaneous initiation of detonations is given by combining equations (\ref{eq:uspfid}), (\ref{eq:uspfid=betaucj}) and (\ref{eq:Tfid}) and setting $\beta = 0.018$, which gives 
\begin{eqnarray}
  \lambda_{\rm crit} &=& 4.6\times 10^5\left(\frac{\beta}{0.018}\right)\left(\frac{\alpha - 1}{20}\right)\left(\frac{\Theta(T_m)}{10^{-3}}\right)\nonumber \\ &&\left(\frac{\Delta T/\To}{1}\right)\left(\frac{X_C}{0.5}\right)^{-2}\rho_7^{-1}\left(\frac{1 + T_o/\To}{2}\right)^{-\alpha}{\rm cm},\label{eq:lambdacrit}
\end{eqnarray}
which given our results for $\beta$ encapsulated in Figure \ref{fig:beta} is good to within a factor of two for our results and that of S09.  Given that the critical radii published by S09 varies by two orders of magnitude and the induction times varies by three orders of magnitude, this gives an excellent approximation.  

\begin{figure}
\includegraphics[width=0.8\textwidth]{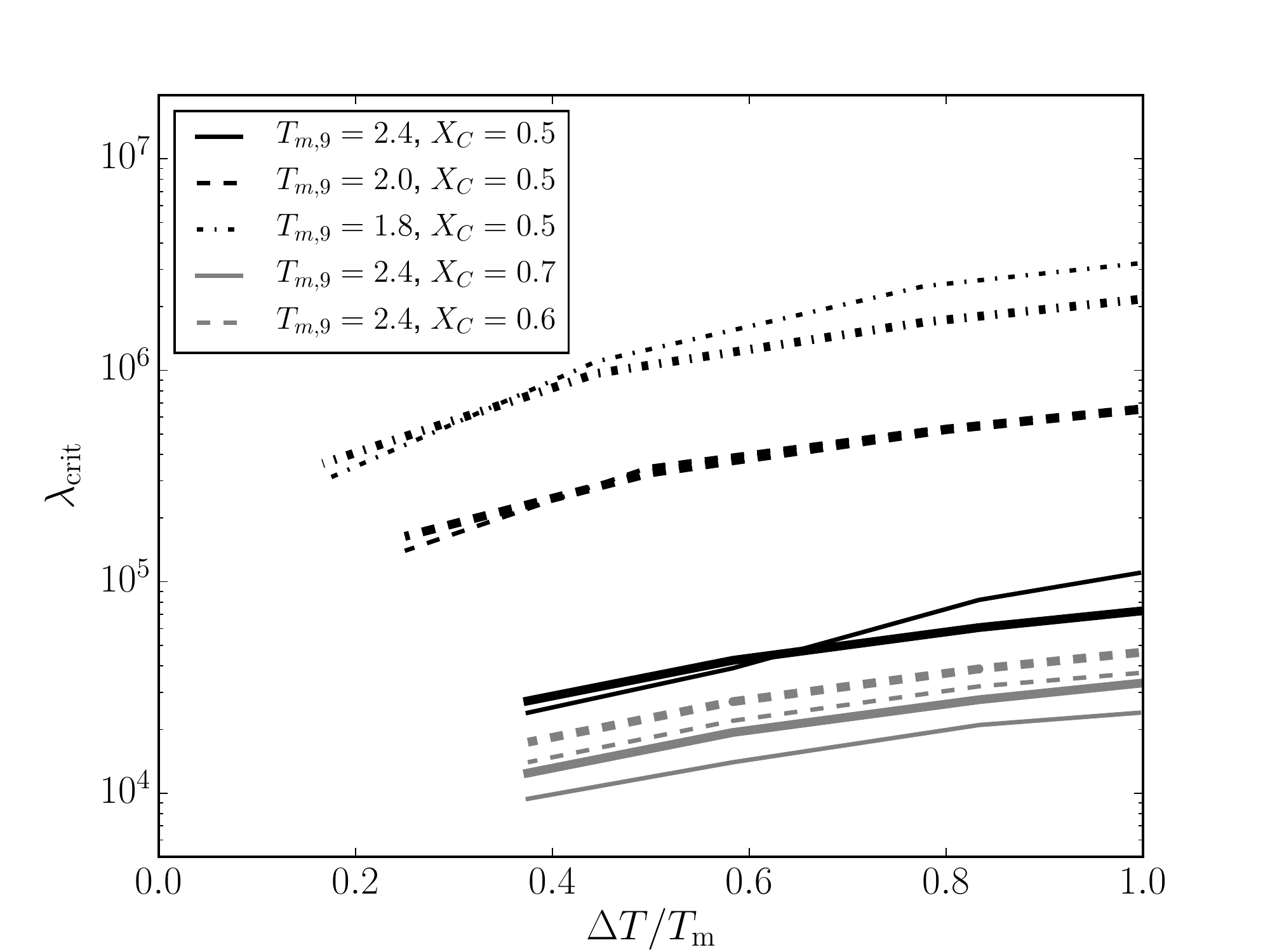}
\caption{The critical radii, $\lambda_{\rm crit}$ from S09 (thin lines) for different $\To$ compared to the semi-analytic expression (corresponding thick lines) given by equation (\ref{eq:lambdacrit}). The analytic expression agrees within a factor of two with the results of S09 for $\To \le 2.4\times 10^9$ K and $X_C \ge 0.5$ over two order of magnitude range in $\lambda_{\rm crit}$.  \label{fig:lambdacrit}}
\end{figure}

Figure \ref{fig:lambdacrit} plots the result of equation (\ref{eq:lambdacrit}; thick lines) against the tabulated results of S09 (thin lines) for $\To \le 2.4\times 10^9$ K and $X_C \ge 0.5$.  The agreement between the semi-analytic condition given by equation (\ref{eq:lambdacrit}) and S09 is excellent in the regime of applicability.  It is also clear from this figure that the scaling of $\lambda_{\rm crit}$ with $X_C$ and $\Delta T/\To$ that we see in equation (\ref{eq:lambdacrit}) holds in S09's tabulated results.  

Finally, we should note that the reported results of S09 come from simulations that were unresolved. Here, S09 performed a resolution study for $T_m = 2.4 \times 10^9$ K and showed that the resolved critical radius is a factor of three larger than the value determined in the unresolved simulations.  Based on this, S09 claimed that the resolved value of the critical radius is likely to be a factor of two or three larger than reported.  In the context of this work, the impact on the fitted value of $\beta$ depends on how the resolved value of the critical radius compared to the reported one.  If the resolved value is universally a factor of two or three higher uniformly, the fitted value of $\beta$ can be increased by the appropriate factor.  However, if the resolved value has a nontrivial behavior compared to the S09 reported values, then a new fit is likely necessary. 
			
\section{Discussion and Conclusions}\label{Conclusions}

In this paper, we have also developed a semi-analytic criterion for spontaneous initiation of detonations by combining the published numerical results of S09 with our analytic expression for $\usp$ (eq. [\ref{usp}] and a condition for detonation captured by equations (\ref{eq:uspfid}) and (\ref{eq:uspfid=betaucj}). This gives an explicit expression for $\lambda_{\rm crit}$ (eq. [\ref{eq:lambdacrit}]),  that fits our results and the results of S09 for $X_C\ge 0.5$ and $\To \le 2.4\times 10^9$ K to within a factor of two.  At smaller $X_C$ and higher $\To$, we suspect that the increasing importance of oxygen burning invalidates our analytic derivation of $\usp$ that assumes pure C burning. Hence our work complements S09 and other works that studied the initiation of detonations numerically.
In addition, the framework developed here provides a means by which spontaneous initiation of detonations in other materials, i.e, helium, can be studied.

We will note that the semi-analytic criterion that we developed expressed by equations (\ref{eq:uspfid}), (\ref{eq:uspfid=betaucj}), and (\ref{eq:Tfid}) depends on two fitted parameters, $\beta \approx 0.018$ and $T_o = 1.5\times 10^9 $ K.  The parameter $\beta$ can be understood as the speed of the fiducial spontaneous wave speed compared to the CJ speed for detonations.  On the other hand, we do not have a explanation or physical understanding for the value of the parameter $T_o$.  

While we have studied the conditions on the hotspot that lead to successful detonation, the generation of these hotspots and their structure demands greater study. The analytic formula derived here suggest that the spontaneous wave speeds at the fiducial temperature is a small fraction of the CJ speed, though it is a good fraction of the CJ speed (and near sonic or supersonic) at the maximum temperature.  How these temperature gradients can be properly preconditioned in realistic white dwarfs remain unclear. Here the analytic criterion for detonations may be useful in studies of hotspot generation and preconditions in burning WDs.

\begin{acknowledgements}
We thank the anonymous reviewer for useful comments.
We acknowledge support the NASA ATP
program through NASA grant NNX13AH43G, and NSF grant AST-1255469.  
\end{acknowledgements}

\bibliographystyle{apj}
\bibliography{ms}

\end{document}